\documentclass[useAMS,usenatbib]{mn2e}
\usepackage{graphicx}
\usepackage{color}
\usepackage{epsfig}
\usepackage{amsmath,amssymb}
\usepackage{natbib}
\usepackage{comment}
\usepackage{multirow}
\usepackage{lscape}
\usepackage{threeparttable}
\usepackage{bm}
\usepackage{ulem}

\def\simge{
    \mathrel{\rlap{\raise 0.511ex
        \hbox{$>$}}{\lower 0.511ex \hbox{$\sim$}}}}
\def\simle{
    \mathrel{\rlap{\raise 0.511ex
        \hbox{$<$}}{\lower 0.511ex \hbox{$\sim$}}}}

\newcommand{\figref}[1]{Fig.~\ref{#1}}
\newcommand{\tabref}[1]{Table~\ref{#1}}
\newcommand{\secref}[1]{Section~\ref{#1}}

\newcommand{\pc}{{\rm pc}}
\newcommand{\msun}{{\rm M}_{\odot}}
\newcommand{\zsun}{Z_{\odot}}
\newcommand{\rsun}{R_{\odot}}

\newcommand{\kyr}{{\rm kyr} }
\newcommand{\Myr}{{\rm Myr} }
\newcommand{\Gyr}{{\rm Gyr} }
\newcommand{\Mpc}{{\rm Mpc} }

\newcommand{\dd}{{\rm d}}

\newcommand{\mc}[1]{\multicolumn{1}{c}{#1}}

\title[Formation of IMBHs in the First Star Clusters]
{Formation of Intermediate-Mass Black Holes through Runaway Collisions in
  the First Star Clusters}

\author[Y. Sakurai et al.]
{Yuya Sakurai$^{1}$\thanks{sakurai@utap.phys.s.u-tokyo.ac.jp}, 
Naoki Yoshida$^{1,2}$, Michiko S. Fujii$^{3}$, Shingo Hirano$^{4}$
\\ \\
$^{1}$Department of Physics, Graduate School of Science, The University of Tokyo, 7-3-1 Hongo, 
Bunkyo-ku, Tokyo 113-0033, Japan\\
$^{2}$Kavli Institute for the Physics and Mathematics of the Universe (WPI), Todai Institute for Advanced Study, The University of Tokyo, \\
~Kashiwa, Chiba 277-8583, Japan \\
$^{3}$Department of Astronomy, Graduate School of Science, The University of Tokyo, 7-3-1 Hongo, Bunkyo-ku, Tokyo 113-0033, Japan\\
$^{4}$Department of Astronomy, The University of Texas, Austin, TX 78712, USA
}
\begin{document}

\date{Draft version \today}

\maketitle

\label{firstpage}

\voffset=-0.4in

\begin{abstract}
  We study the formation of massive black holes in the first star clusters.
  We first locate star-forming gas clouds in proto-galactic haloes of $\gtrsim \!10^7\,\msun$
  in cosmological hydrodynamics simulations and use them to generate
  the initial conditions for star clusters with masses of $\sim \!10^5\,\msun$.
  We then perform a series of direct-tree hybrid $N$-body simulations to
  follow runaway stellar collisions in the dense star clusters.
  In all the cluster models except one, runaway collisions occur
  within a few million years, and the mass of the central, most massive star
  reaches $\sim \!400-1900\,\msun$.
  Such very massive stars collapse to leave intermediate-mass black holes (IMBHs).
  The diversity of the final masses may be attributed to the differences
  in a few basic properties of the host haloes such as mass, central gas velocity
  dispersion, and mean gas density of the central core.
  Finally, we derive the IMBH mass to cluster mass ratios, and compare them with
  the observed black hole to bulge mass ratios in the present-day Universe.
\end{abstract}

\begin{keywords}
intermediate mass black holes -- galaxies: star clusters -- stellar dynamics
\end{keywords}

\section{Introduction}
\label{sec:introduction}
The origin of supermassive black holes (SMBHs) at $z\gtrsim \!6$ remains largely
unknown \citep[e.g.][]{Mortlock2011,Wu2015}. 
Prompt formation of such massive objects appears implausible, and
thus viable models posits formation of black hole `seeds'
and also their efficient growth. There are a few suggested formation mechanisms
of the seed black holes.
For example, very massive Population III stars, which are thought to be formed
in primordial gas clouds \citep[e.g.][]{Susa2014,Hirano14}, 
leave remnant BHs with masses greater than $\sim \! 100\,\msun$.
While the Population III remnants appear as promising BH seeds,
it is unlikely that the BH subsequently
grow in mass efficiently to be as massive as the observed SMBHs by $z= \! 6-7$
\citep[e.g.][]{Madau_Rees2001,Haiman_Loeb2001,Schneider2002,Volonteri2003}.
Another popular physical model is the so-called direct-collapse of super-massive stars
(e.g. \citealt{Loeb_Rasio1994,Oh_Haiman2002,Bromm2003,Begelman2006}).
%
The model is motivated to kick-start the formation of a SMBH from 
a BH with $\sim \!10^5\,\msun$
at a sufficiently
early epoch. The direct-collapse model, however, often resorts to a few peculiar
conditions such as the existence of luminous galaxies nearby, and hence the overall
formation rate of massive BHs remains rather uncertain.

There is yet another mechanism of the formation of massive black holes
in the early universe.
It has been suggested that some star clusters in the present-day Universe
harbor intermediate mass black holes (IMBHs) at their centres.
Such IMBHs, if they exist, are likely remnants of very massive stars
that could have been formed through runaway stellar collisions
(\citealt{Ebisuzaki01}; \citealt{Vanbeveren09}; \citealt{Devecchi10,Katz15}).
In the context of early structure formation,
\citet{Omukai08} consider dense star clusters formed in the so-called
atomic-cooling haloes with virial temperature of $\gtrsim \!10^4\,{\rm K}$. 
They argue that a metal-enriched gas cloud with metallicity greater
than $Z_{\rm cr}\sim \!5\times10^{-6}\,\zsun$
can undergo dust-induced cooling and fragmentation, to produce a star cluster.
\citet{Devecchi10} study formation of very massive stars and BH seeds 
based on a hierarchical galaxy formation model.
Their semi-analytic model predicts a large BH mass density of
$\rho_{\rm seed}\sim \!2000-4000\,\msun\,\Mpc^{-3}$ at $z\gtrsim \!6$.

Star cluster formation at high redshift has been also studied
by numerical simulations \citep{Boley09,Trenti15}.
\citet{Bromm_Clarke02} perform SPH simulations of early dwarf galaxy formation
and show that globular clusters of $\sim \!10^5\,\msun$ are formed in small-mass haloes.
\citet{Kimm16} perform cosmological radiation-hydrodynamics simulations of 
globular cluster formation at $z>10$. 
They find that gas fragmentation induced by metal cooling 
drives the formation of dense star clusters of 
$\sim \!6\times10^5\,\msun$ with half-light radius $\lesssim \! 1\,\pc$
in about ten million years.

Overall, the formation of dense star clusters in the early universe
appears plausible, but there are surprisingly few simulations to date
that follow the evolution of a star cluster directly.
\citet{Katz15} use a cosmological simulation to
study the dynamical evolution of a star cluster in a metal-enriched mini-halo.
With the aid of direct $N$-body simulations,
they show that runaway stellar collisions occur in the cluster,
to yield a very massive star with mass of
$\sim \! 300-1000\,\msun$. The result provide an interesting formation path
of an IMBH. Unfortunately, the cluster model of \citet{Katz15} is
based on only one specific mini-halo, and thus it remains unclear how rare
is such IMBH formation in the early universe.

In this study, we use direct-tree hybrid $N$-body simulations
to follow the evolution of a number of
dense star clusters. We generate realistic initial conditions by utilizing
the output of cosmological hydrodynamics simulations. 
We show that all reasonable cluster models except one yield a very massive
star of $\gtrsim \! 1000\,\msun$ 
as a product of runaway collisions in a few million years.

The rest of the present paper is organised as follows.
In Section 2, we describe our cosmological simulations and direct-tree hybrid N-body simulations. 
There, we also explain generation of the initial conditions from the cosmological simulations.
In Section 3, we show the results of the hybrid $N$-body simulations.
Finally in Section 4, we give summary and concluding remarks.

\section{Numerical Methods}
\label{sec:method}
Our simulations are performed in the following three-step manner.
We first run cosmological hydrodynamics simulations of early galaxy formation.
We use the output to locate a number of proto-galactic haloes
that host a star-forming gas cloud.
We then replace the gas cloud with a dense star cluster on the assumption
that numerous stars with a wide range of masses are formed in the cloud.
The positions and the velocities of the
stars are configured by adopting a few simple models.
The realizations generated in this manner serve as the
initial conditions for star cluster evolution simulations.
Finally, we run direct-tree hybrid $N$-body simulations
and follow the stellar dynamics.

\subsection{Cosmological simulations}
We use the parallel $N$-body/Smoothed Particle Hydrodynamics (SPH) code Gadget-2
\citep[][]{Springel05}, suitably modified as in \citet{Hirano14} so that the
formation of primordial gas clouds can be followed
(see also \citealt{yoshida03,yoshida08}).
The simulations are initialized at $z_{\rm ini}= \! 99$ with a box size $10\,h^{-1}\,\Mpc$
using the MUSIC software \citep{Hahn11}. 
Cosmological parameters are adopted from the latest Planck data \citep[][last column of their table 4]{PlanckCollaboration15}.
The box size is chosen to be sufficiently large to locate
about 10 haloes with virial mass $ \sim \! 10^7-10^8\,\msun$
at redshift $z= \! 10-20$ \citep{Reed07}.
We first run a dark matter only simulation with $N= \! 512^3$,
and run a friends-of-friends halo finder to identify dark haloes at $z= \! 10$.
Next, we perform zoom-in simulations for the selected target haloes
with a high spatial resolution.
With the multi-level zoom-in technique, we achieve a mass resolution
of $m_{\rm DM}\sim \! 1\,\msun$.
The resolution is designated by considering that, in our star cluster
simulations (\secref{sec:Nbody}), 
the DM particle mass should be smaller than the minimum stellar
mass that is $3\,\msun$ in our fiducial models
(\secref{sec:generation}).
We perform the zoom-in simulations including SPH particles but 
with radiative cooling other than atomic hydrogen cooling switched off.
Molecular hydrogen cooling is disabled in order to prevent 
gas cloud formation in early mini-haloes.
The SPH simulations are stopped when the target haloes collapse gravitationally,
and the central gas density
reaches $n_{\rm H}\sim \! 10^7\,{\rm cm^{-3}}$.
We run simulations for a total of eight haloes. The basic halo properties are listed
in \tabref{tab:SC}.


\begin{table*}
  \begin{center}
    \caption{
      Properties of the host haloes and the main results of our star cluster simulations
      with a fiducial set of model parameters, $\alpha_{\rm sfe}= \! 6.32\times10^{-4}$, $m_{\rm min}= \! 3\,\msun$, $m_{\rm max}= \! 100\,\msun$, 
$\beta= \! 2.35$, $Q= \! 0.5$ and $m_{\rm DM}= \! 1.87\,\msun$. 
      Overlines indicate that the values are averaged over 3 realizations.
    }    
      \begin{tabular}{lrrrrrrrrrrrrrrr} \hline
       &  \mc{z}  &  \mc{$R_{\rm vir}$} &  \mc{$M_{\rm vir}$} & \mc{$\overline{M}_{\rm cl}$} & \mc{$\overline{N}$}   & \mc{$\overline{r}_{\rm c}$} & \mc{$\overline{\rho}_{\rm c}$} & \mc{$\overline{t}_{\rm rh}$}  & \mc{$\overline{t}_{\rm rc}$}  & \mc{$\overline{\epsilon}_{\rm sfe}$} & \mc{$M_{\rm DM}$} & \mc{$N_{\rm DM}$} & \mc{$\overline{m}_{\rm max,f}$} & \mc{$\overline{N}_{\rm coll}$}  \\
        &              &  \mc{($\pc$)}        &  \mc{($10^7\,\msun$)}& \mc{($10^4\,\msun$)}         & \mc{($10^3$)} & \mc{($\pc$)}                        & \mc{($\msun\,{\rm pc^{-3}}$)}  & \mc{($\Myr$)}                                & \mc{($\kyr$)}                               &  \mc{(\%)}                                           & \mc{($10^7\,\msun$)} & \mc{($10^7$)}          & \mc{($\msun$)}                           &   \\
\hline \hline
A & 19.7  & 281  & 4.03 &   16.4 &   19.9 &   0.401 &  $6.45\times10^5$ &  19.7 & 528  & 5.91  & 4.79 & 2.56 &   929 &   11.7  \\ 
B & 19.6  & 276  & 2.97 &   13.0 &   15.7 &  0.387 &   $5.82\times10^5$ &  12.6 & 783  & 6.12  & 3.78 & 2.02 &   409 &   4.67  \\ 
C & 19.7  & 208  & 2.03 &   12.1 &   14.7 &  0.380 &   $8.36\times10^5$ &  9.67 & 67.1  & 10.1    & 6.60 & 3.53 &   1330 &   18.3  \\ 
D &  14.9 & 321 & 2.60 &   11.7 &   14.1 &   0.357 &   $9.75\times10^5$ &  13.6 & 8.93  &  7.16  & 5.67 & 3.03 &   971  &   13.7   \\ 
E & 17.1  & 264  & 1.47 &  4.76 &  5.76 &    0.224 &   $1.16\times10^6$ &  4.42 & 2.98 &  8.15 & 3.25 & 1.74  &   773  &   9.67   \\ 
F &  16.5  & 312 & 2.01 &  9.00 &   10.8 &   0.662 &   $7.05\times10^5$ &  15.0 & 3.62  &  8.67  & 5.13 & 2.74 &   1100  &   14.0   \\ 
G & 16.9  & 242 &1.99 &   12.5 &   15.0 &   0.353 &   $1.01\times10^6$  &  10.1 & 4.25  & 9.48  & 4.17 & 2.23  &   1660 &   25.0  \\ 
H & 11.7   & 541 &4.22  &  7.70 &  9.32 &    0.276 &   $1.08\times10^6$  &  10.8 & 0.807 & 5.55 & 5.25 & 2.81 &   964 &   15.0   \\ 

\hline

    \end{tabular}
    \label{tab:SC}
    \begin{tablenotes}
    
      \small
    \item {\bf Notes 1: }
      Properties of the haloes when the central gas density is
      $n_{\rm H}= \! 10^7\,{\rm cm^{-3}}$; 
      Column 2: redshift, 
    Column 3: virial radius, and 
    Column 4: virial mass.
      \item {\bf Notes 2: } Generated cluster models; 
    Column 5: total star mass, 
    Column 6: total number of star particles, 
    Column 7: core radius,
    Column 8: core density,
    Column 9: half-mass relaxation time (equation \ref{eq:t_rh}),
    Column 10: central relaxation time (equation \ref{eq:trc}),
    Column 11: global star formation efficiency $\epsilon_{\rm sfe}\equiv M_{\rm cl}/M_{\rm gas}(<R_{\rm cl})$ where 
    $M_{\rm gas}(<r)$ is enclosed mass of gas in the original halo data, 
    Column 12: total DM mass, and
    Column 13: total number of DM particles.
    The core radius and core density are calculated following \citet{Casertano85}. 
    The values $r_{\rm c}$, $\rho_{\rm c}$ and $t_{\rm rh}$ are computed
    using bound star particles.
  \item {\bf Notes 3: } Results of the hybrid N-body simulations;
    Column 14: maximum stellar mass formed via runaway collision, and
    Column 15: average number of collisions during the simulations. 
    
    \end{tablenotes}
  \end{center}
\end{table*}

\subsection{Generation of star cluster/DM distributions}
\label{sec:generation}
We generate the initial conditions for our stellar dynamics simulations
directly from the outputs of cosmological zoom-in simulations.
For each target halo, we dump a snapshot when the central
gas density is $n_{\rm H}\sim \!  10^7\,{\rm cm^{-3}}$.
The density roughly corresponds to a critical density for fragmentation
for a gas cloud with metallicity of $Z\sim \!  10^{-4}\,\zsun$ (see fig. 5 of \citealt{Omukai08}).
We expect the cloud is already gravitationally unstable, 
and likely yields multiple stars. However, our zoom-in simulations do not
resolve the formation of individual stars, and thus we employ the following
simplified model to place stars within the parent gas cloud.

We sample a fraction of the SPH particles as ``stars''. A selected star
particle is re-assigned its mass and velocities, while its position is
kept the same as that of the original SPH particle.
We use the following five {\it physical} parameters to determine the sample
probability and to calculate the stellar mass and velocities: 
local star formation efficiency (SFE) $\alpha_{\rm sfe}$, 
the minimum stellar mass $m_{\rm min}$, the maximum stellar mass $m_{\rm max}$,
index $\beta$ of a power-law initial mass function (IMF) $\dd N/\dd m\propto m^{-\beta}$,
and virial ratio $Q$ which is the ratio of the total stellar kinetic energy 
to the total stellar potential energy.
The probability of replacing a SPH particle $i$ with a
star particle is calculated
according to the local SFE \citep{Fujii15} given by 
\begin{equation}
\epsilon_{{\rm loc},i}= \! \max\left(\alpha_{\rm sfe}\sqrt{\frac{n_{{\rm H},i}}{1\,{\rm cm^{-3}}}}{\rm e}^{-r_i/R_{\rm cl}},1.0\right)
\times\frac{m_{{\rm gas},i}}{\overline{m}_{\rm s}}, \label{eq:sfe}
\end{equation}
where $r_i$ is the distance of the SPH particle from the point of maximum density
(cloud centre),  $R_{\rm cl}$ is the radius where the enclosed gas mass is equal to
that of DM,
$m_{{\rm gas},i}$ is SPH particle mass and $\overline{m}_{\rm s}$ is the mean
stellar mass for the specified IMF.
The equation \eqref{eq:sfe} is based on the star-formation law
$\dot{\rho}_{\rm star}\propto t_{\rm ff}^{-1}$
\citep{Schmidt59,Kennicutt98}, where the star formation rate
$\dot{\rho}_{\rm star}$ is inversely proportional
to free-fall time $t_{\rm ff}\propto n_{\rm H}^{-1/2}$.
The factor $m_{{\rm gas},i}/\overline{m}_{\rm s}$ guarantees mass 
conservation when the gas to star conversion is performed.
The exponential cutoff is applied to set a finite cluster size,
but the choice of the value of $R_{\rm cl}$ is unimportant.
To assign the velocity to each star particle, 
we re-scale the velocity of the SPH particle as
\begin{equation}
\bm{v}_{\rm star}= \! \sqrt{\frac{Q}{T/|W|}}(\bm{v}_{\rm SPH}-\bm{\overline{v}}), \label{eq:vel}
\end{equation}
where $\bm{v}_{\rm SPH}$ is the SPH particle velocity after the cloud's bulk velocity
is subtracted, 
$\bm{\overline{v}}= \! \sum_{\rm replaced}m_{\rm star}\bm{v}_{\rm SPH}/\sum m_{\rm star}$, $m_{\rm star}$ is stellar mass,
$T= \! \sum_{\rm replaced}m_{\rm star}\bm{v}_{\rm SPH}^2/2$,
and $W$ is potential energy of the stars.
With the fiducial virial ratio of $Q= \! 0.5$, we can realize a marginally
stable cluster.
For each sample, we generate three realizations using different random number seeds
to select star particles. We thus run a total of 24 simulations
for our fiducial model. We further run additional simulations
to examine the effect of model parameters, as will be discussed in 
\secref{sec:parameter}.

The dark matter distribution is kept essentially the same as in the original
cosmological simulation.
In practice, we split DM particles so that all the DM particles have the
same mass $m_{\rm DM}$ which is chosen as the minimum dark matter particle mass
in the cosmological simulation.
The splitting is necessary to avoid artificial mass segregation of DM particles
in the hybrid $N$-body
simulations (\secref{sec:Nbody}).
When the splitting is done, the daughter particles are randomly distributed
in a sphere whose radius is the mean separation of the DM particle, 
and retain the same velocity.
The DM halo's bulk velocity is also subtracted, and the particle velocities
are re-scaled so that the DM virial ratio
becomes 0.5, as in equation \eqref{eq:vel}.
By doing this, we prevent the outer part of the DM halo from 
evaporating during our star cluster simulations.

The fiducial parameters for the initial condition generation are set as
$\alpha_{\rm sfe}= \! 6.32\times10^{-4}$, $m_{\rm min}= \! 3\,\msun$, $m_{\rm max}= \! 100\,\msun$, 
$\beta= \! 2.35$, $Q= \! 0.5$ and $m_{\rm DM}= \! 1.87\,\msun$.
The resulting star cluster/DM global properties are shown in 
\tabref{tab:SC}.
The value of $\alpha_{\rm sfe}$ is chosen such that the particle number
in Model A is about $\sim \!  2\times10^4$.
With this choice, the global star formation efficiency 
(column 10 in \tabref{tab:SC}) is 
$\epsilon_{\rm sfe}\sim \!  0.06$-$0.1$,
which is approximately consistent with the value of $\sim \! 0.1$
found in the hydrodynamics simulations of \citet{Kimm16}.
Note that $Q$ is set to 0.5, but it does not necessarily mean that the system
is in virial equilibrium since not all the star particles are bound.
The Salpeter mass function is assumed throughout the present paper.
We note here that stars with lower masses of $m_{\rm min} < 1 \,\msun$
may exist in real clusters.  
Setting a smaller $m_{\rm min}$ makes the number of star particles very large
and our hybrid $N$-body simulations
become virtually impractical. We study the effect of varying
$m_{\rm min}$, as well as the other parameters, in \secref{sec:parameter}.

\subsection{Direct-tree hybrid N-body simulations}
\label{sec:Nbody}
The stellar dynamics simulations are performed using the hybrid $N$-body
code BRIDGE \citep{Fujii07}.
The code follow the orbits of the star particles by a direct method
in a dynamically consistent
manner with DM particles whose motions are calculated 
by a tree method.
In the current version, the sixth-order Hermite integrator is applied for the direct integration \citep{Nitadori08}.
The NINJA scheme realizes better parallel efficiency \citep{Nitadori06}.
The phantom-GRAPE library is used \citep{Tanikawa13} to speed-up the 
gravitational force calculation.

A pair of stars can collide and merge when its separation 
$d\equiv|\bm{r}_1-\bm{r}_2|$ becomes less than the sum of the stellar radii 
$R_{*,1}+R_{*,2}$, i.e., $d<R_{*,1}+R_{*,2}$.
This merger criterion is well tested by \citet{Gaburov10}, who show that 
  the criterion gives sufficiently accurate results when compared to the results of direct 
  hydrodynamics simulations of stellar three-body interactions.

The stellar radius in our simulations is given by the fitting formula of \citet{Tout96}
for the non-evolving zero-age main-sequence stars with $Z= \! 0.02$.
The fitting, which is valid for stellar mass smaller than $100\,\msun$, is extrapolated to 
larger stellar mass and thus it might possibly cause underestimation of the radii, especially for very massive stars.
For example, the radii of stars with mass $100, 200, 500, 1000\,\msun$ 
from the formula of \citet{Tout96} are $17, 28, 54, 87\,\rsun$ respectively, 
while those from the interior structure calculations of \citet{Ishii99} for massive stars are 
$18, 40, 160, 3000\,\rsun$ respectively
and those of \citet{Yungelson08} are $14, 27, 66, 129\, \rsun$ respectively.
Note that the stellar radius is generally smaller for stars with
lower metallicity \citep[e.g. ][]{Baraffe91,Baraffe01}.
Although we could technically construct a model of stellar radii for $Z= \! 10^{-4}\,Z_\odot$ as in 
\citet{Katz15}, we have decided to use Tout's fitting formula of the solar metallicity for simplicity.
We study the effect of adopting a different model of stellar
radius in (\secref{sec:parameter}).

\begin{table}
  \begin{center}
    \caption{Parameters of the hybrid N-body simulations.}
    \begin{tabular}{lrrrrrrr} \hline
       \mc{$\eta$}  & \mc{$\Delta t$} & \mc{$L_{\rm box}$} & \mc{$n_{\rm crit}$} & \mc{$\theta$} & \mc{$\epsilon_{\rm cl}$} & \mc{$\epsilon_{\rm DM}$} \\
         & \mc{(yr)} & \mc{(pc)}  &  & & (pc) & (pc) \\ \hline \hline
       0.11     & $1.16\times 10^3$       & 1024/2048         & 512 & 0.5 & 0 & 0.0313  \\ \hline
    \end{tabular}
    \label{tab:simparam}
  \end{center}
\end{table}

Parameters of the $N$-body simulations are presented in \tabref{tab:simparam}, 
where $\eta$ is an integration accuracy parameter (equation 16 in \citealt{Nitadori08}), 
$\Delta t$ is a tree timestep, 
$L_{\rm box}$ is a box size of the simulations, 
$n_{\rm crit}$ is the maximum group size for GRAPE calculation \citep{Makino91},
$\theta$ is the tree opening angle, and
$\epsilon_{\rm cl}$ and $\epsilon_{\rm DM}$ are softening
lengths for star clusters and DM, respectively.
The accuracy parameter $\eta$ is chosen so that the total energy
errors during the simulations do not exceed $0.04\%$.
The simulation boxsize $L_{\rm box}$ is set to be 1024 pc or 2048 pc,
for all the particles to be encompassed during the simulations.
Convergence of the results has been checked for $\Delta t$, $L_{\rm box}$
and $\epsilon_{\rm DM}$.
The simulations are stopped at $t= \!  3\,\Myr$, when the central massive
star is supposed to end its life.

\section{Results}
\label{sec:results}
\begin{figure*}
    \centering
    \includegraphics[width=1.0\textwidth]{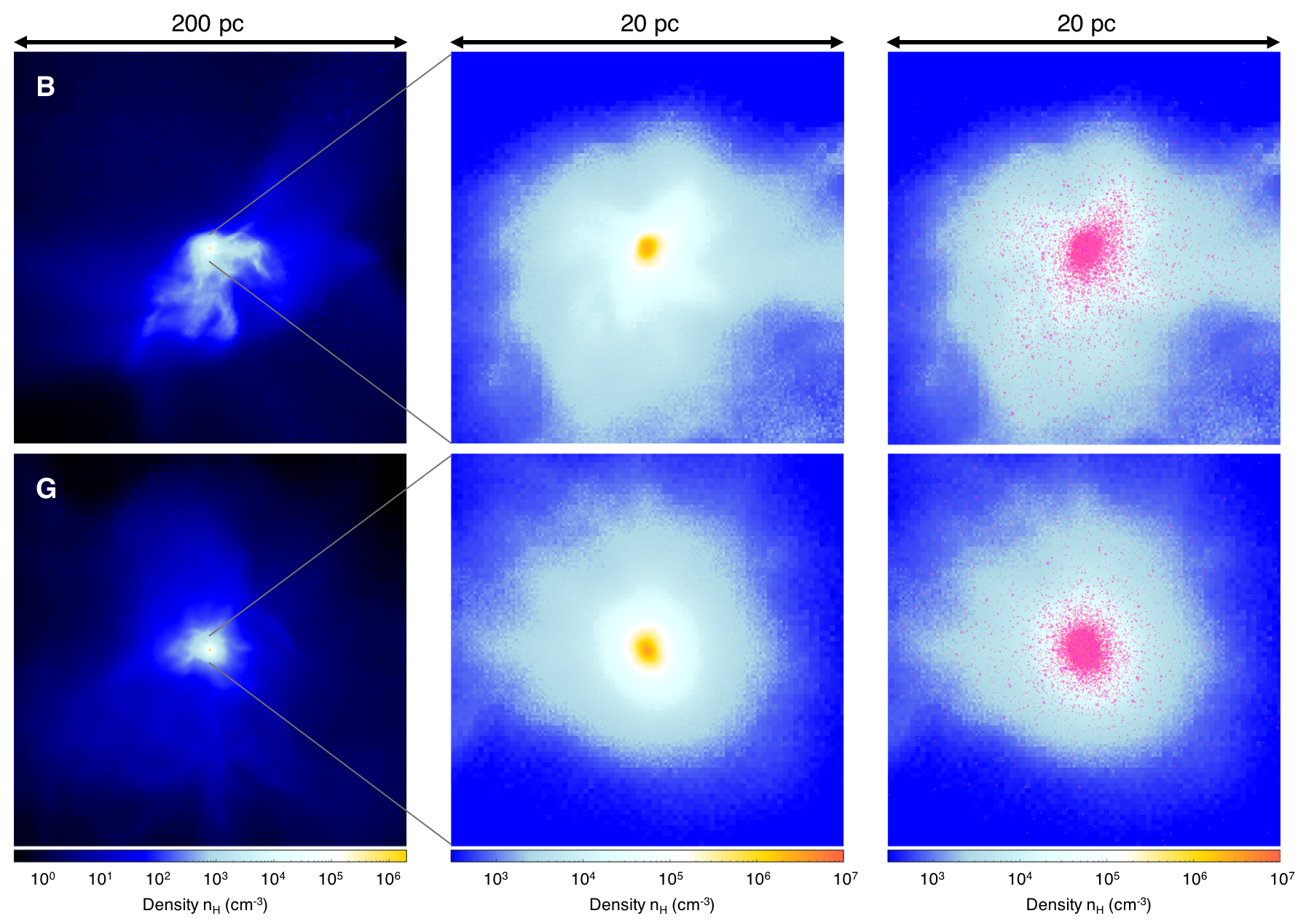}
    \caption{
      We plot the projected gas density distribution for halo 
      B (top panels) and halo G (bottom panels). 
    The size of the region is $200\,\pc$ (left) and $20\,\pc$ (middle). 
    We also compare the generated star distribution (magenta dots) on the right.
    }
    \label{fig:pos_gas}
\end{figure*}

The properties of the eight haloes are summarized in \tabref{tab:SC}.
The virial masses are $\sim \! (1.5$ - $4)\times10^7\,\msun$, and
the formation epochs are $z = \!  11 - 20$.
\figref{fig:pos_gas} shows the gas distribution for our
Halo B (top panels) and G (bottom panels), respectively.
Halo B shows the features of turbulent motions in the
gas distribution, whereas Halo G appears dynamically relaxed, and
is approximately spherical. The generated stellar distributions
are also shown in the right panels of \figref{fig:pos_gas}. 
Most of the stars are located within the central few parsec region but 
a small number of stars are also found in the outer region.

\subsection{Fiducial models}
\label{sec:fiducial}
With our fiducial model parameters, the initial stellar
mass in the cluster ranges from $\sim \!  5\times10^4\,\msun$
to $\sim \!  1.6\times10^5\,\msun$
(see \tabref{tab:SC}).

\begin{figure}
    \centering
    \includegraphics[width=0.5\textwidth]{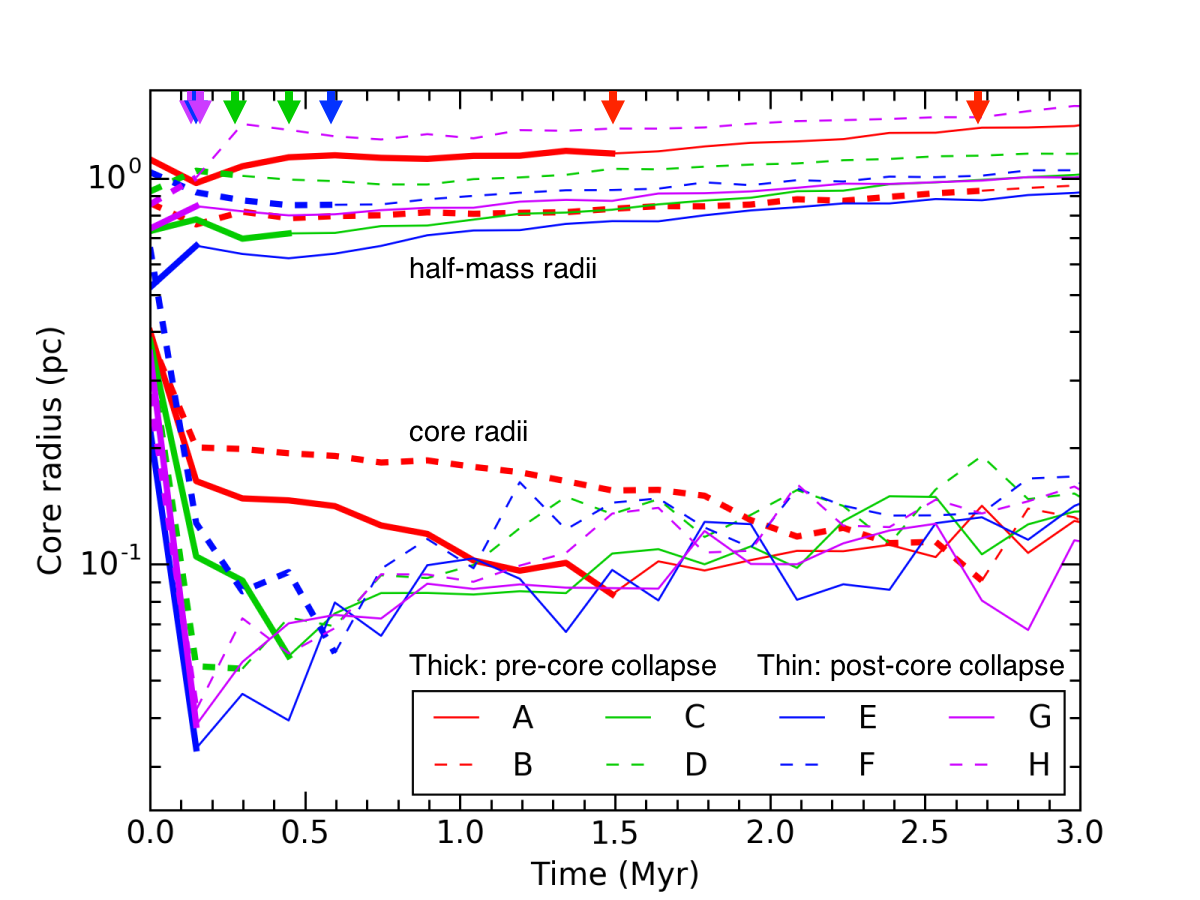}
    \caption{
    Evolution of the core radius and the half-mass radius for the 8 models A to H.
    The thick lines indicate pre-core collapse phases while the thin
    lines indicate post-core collapse phases.
    The arrows mark the core collapse time when the core radius is smallest.
    }
    \label{fig:r_core}
\end{figure}

\figref{fig:r_core} shows the time evolution of the cluster core radii.
We calculate the core radii following the procedure of \citet{Casertano85}.
The core shrinks quickly within about a million years
because of the short crossing time 
$t_{\rm cr}\equiv r_{\rm h}/\sigma\lesssim \! 0.1\,\Myr$ of the cluster systems,
where the half-mass radius $r_{\rm h}$ is typically 
$\sim \!  1\,\pc$ and three-dimensional velocity dispersion $\sigma$
is $\sim \! 10\,{\rm km\,s^{-1}}$.

\figref{fig:r_core} also shows the evolution of $r_{\rm h}$.
In contrast to that of the core radii, $r_{\rm h}$ remains roughly constant with time.
This suggests that the runaway collision is initiated by core collapse
and not driven by cold collapse of the whole cluster due to
the initially non-equilibrium configuration.

Core collapse time $t_{\rm cc}$, as defined by the time when the core radius has its
minimum, ranges from $\lesssim \! 0.1\,\Myr$ to $\sim \! 2.7\,\Myr$ for our 24 simulations.
Rewriting in terms of the half-mass relaxation time $t_{\rm rh}$ (\tabref{tab:SC}) given by
\begin{equation}
t_{\rm rh}= \! \frac{0.651\,{\rm Gyr}}{\ln(\gamma N)}
\frac{1\,\msun}{\overline{m}_{\rm s}}
\left(\frac{M_{\rm cl}}{10^5\,\msun}\right)^{1/2}
\left(\frac{r_{\rm h}}{1\,{\rm pc}}\right)^{3/2}, \label{eq:t_rh}
\end{equation}
where $M_{\rm cl}$ is cluster mass and $\gamma$ is $\sim \! 0.015$ 
in a system with a wide mass spectrum \citep{Giersz96,Gurkan04},
we find that $t_{\rm cc}$ ranges from $\lesssim \! 0.01t_{\rm rh}$ to $\sim \! 0.2t_{\rm rh}$.

The core collapse time $t_{\rm cc}$ is expected to be
proportional to the central relaxation time-scale
\citep[][see \tabref{tab:SC}]{Gurkan04,Fujii14}, which is
\begin{equation}
t_{\rm rc}= \! \frac{0.065\sigma_{\rm c}^3}{G^2\overline{m}_{\rm s}\rho_{\rm c}\ln(\gamma N)}, \label{eq:trc}
\end{equation}
where $\sigma_{\rm c}$ is the three-dimensional velocity dispersion 
at a cluster centre, which we calculate using central 20 stars.
The time $t_{\rm cc}$ ranges from $\sim \!  2 t_{\rm rc}$ to $200 t_{\rm rc}$.
Note that, in contrast to the result of \citet{Fujii14}, we find 
significant scatter in the ratio of $t_{\rm cc}/t_{\rm rc}$
even though $m_{\rm max}/\overline{m}_{\rm s}$ is fixed. This is likely
due to the variation in the compactness of the initial 
stellar distribution.

\begin{figure}
    \centering
    \includegraphics[width=0.5\textwidth]{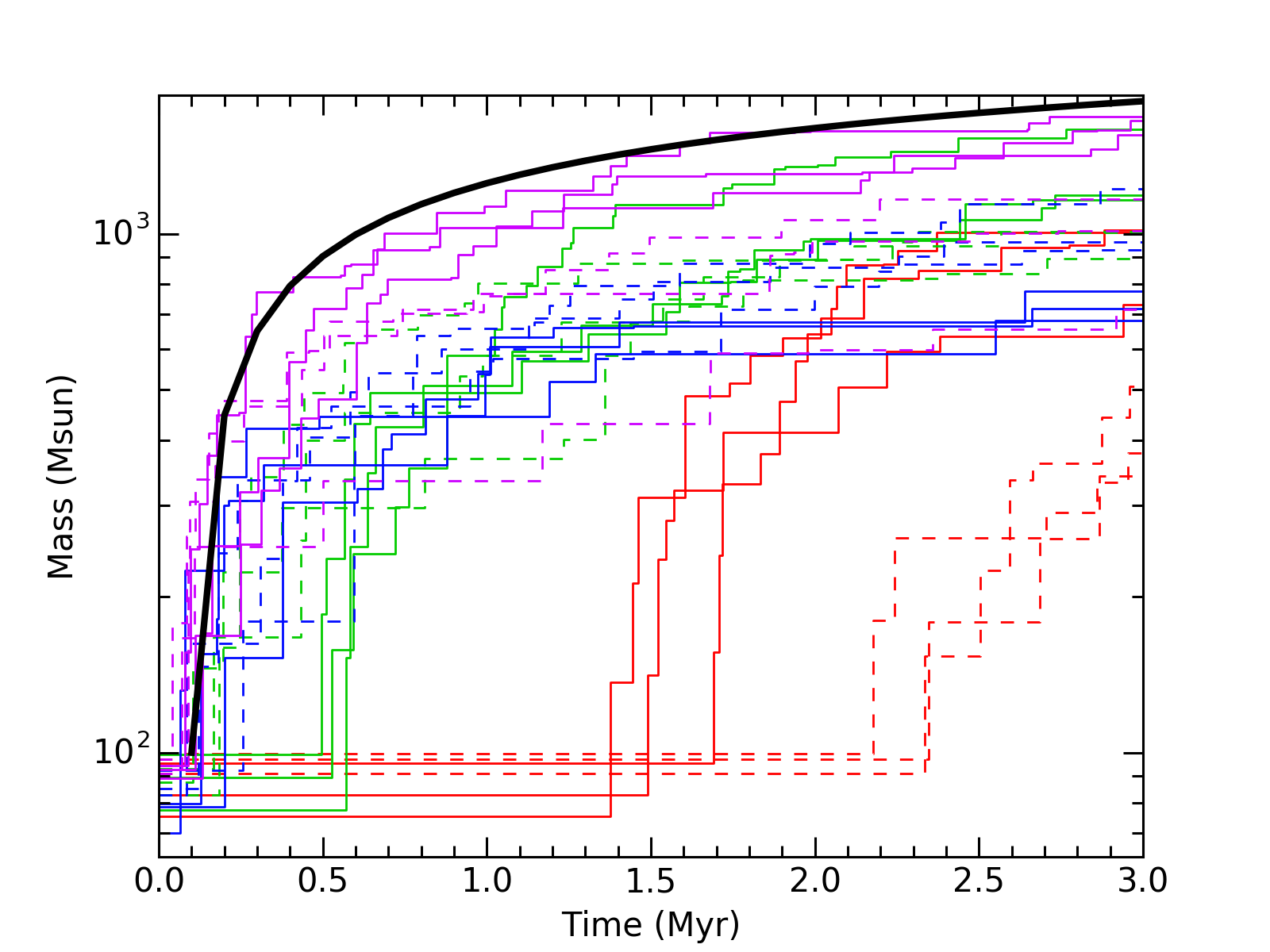}
    \caption{
    Mass evolution of the stars which undergo runaway collision at the cluster centre.
    The analytical expression of equation \eqref{eq:massev} is also plotted
    (black thick line) for the model G
    with $m_{\rm seed}= \! 100\,\msun$, $M_{\rm cl}= \! 1.25\times10^5\,\msun$, $t_{\rm cc}= \! 0.1\,\Myr$, and $f_{\rm c}\ln\Lambda= \! 1$.
    The line types and colours are the same as in \figref{fig:r_core}.}
    \label{fig:mass_ind}
\end{figure}

The final stellar mass $m_{\rm max,f}$ and the number of collisions
$N_{\rm coll}$ by $t = \!  3\,\Myr$ are listed in \tabref{tab:SC}.
In all our runs, runaway collision occurs with $N_{\rm coll}\sim \! 5$ - $25$ and 
the final stellar mass exceeds the threshold mass
$\sim \! 300\,\msun$ for gravitational collapse (\citealt{Heger03}; \citealt{Yoon12,Spera17});
IMBHs are likely to be left in the clusters.
\figref{fig:mass_ind} shows the mass of the central star that undergoes
runaway collisions.
We also show the mass evolution calculated based on the analytic model of
\citep{PortegiesZwart02}:
\begin{equation}
m= \! m_{\rm seed}+4\times10^{-3}M_{\rm cl}f_{\rm c}\ln\Lambda\ln\left(\frac{t}{t_{\rm cc}}\right) \label{eq:massev},
\end{equation}
where $m_{\rm seed}$ is the seed mass of a star which commences runaway process,
$f_{\rm c}$ is the fraction of binaries that contribute to collisions, 
$\ln\Lambda$ is the Coulomb logarithm, and $t_{\rm cc}$ is the core collapse time.
The equation is derived by integrating the mass growth rate which is essentially
the average mass increase per collision times
the mean collision rate of binaries.
The rate of mass accumulation through collisions in our simulations
is well described by this model.

\subsection{Dependence on model parameters}
\label{sec:parameter}

\begin{table*}
  \begin{center}
    \caption{
    Additional star cluster models with different parameters. 
    The main difference from our fiducial model is described in the last column.
    The values of $M_{\rm DM}$ and $N_{\rm DM}$, if DM is included, are the same as
    in the corresponding fiducial model and thus are not listed here.
    Note that the initial distribution of the stars for Model 
    AnoDM and Arad are exactly the same as in the fiducial model A.
    Unless mentioned, the values are obtained by averaging over 3 realizations.
    }
      \begin{tabular}{lrrrrrrrrrrl} \hline
      & \mc{$\overline{M}_{\rm cl,4}$} & \mc{$\overline{N}_{\rm 3}$}   & \mc{$\overline{r}_{\rm c}$} & \mc{$\overline{\rho}_{\rm c,5}$} & \mc{$\overline{t}_{\rm rh}$} & \mc{$\overline{t}_{\rm rc}$} & \mc{$\overline{\epsilon}_{\rm sfe}$} & \mc{$\overline{m}_{\rm max,f}$} & \mc{$\overline{N}_{\rm coll}$} & \mc{Notes} \\
      & \mc{($10^4\,\msun$)}               & \mc{($10^3$)}                         & \mc{($\pc$)}                      & \mc{($\msun\,{\rm pc^{-3}}$)}     & \mc{($\Myr$)}                         & \mc{($\kyr$)}                     & \mc{(\%)}                                           & \mc{($\msun$)}                           &                                                 & \\
\hline \hline
A          &   16.4 &   19.9 &    0.401 & $6.45\times10^5$  & 19.7 & 528 & 5.91   & 929 & 11.7   & fiducial model (\tabref{tab:SC}) \\  \hline
AnoDM &   16.4 &   19.9 &    0.401 & $6.45\times10^5$ & 19.7 & 528 & 5.91   & 915 & 10.7   & no DM \\ 
Arad     &   16.4 &   20.0 &    0.400 & $6.30\times10^5$ & 19.5 & 553 & 5.90   & 958  & 9.00   & half radii for stars, 1 realization  \\ 
Amax    &   16.5 &   18.6 &    0.394 & $7.31\times10^5$ & 17.2 & 157 & 5.95   &1510 &   11.0 & $m_{\rm max}= \! 200\,\msun$, 1 realization  \\ 
Amin     &   16.6 &   53.3 &    0.401 & $7.94\times10^5$ & 44.6 & 1040 & 5.99   & 980  &  12.0 & $m_{\rm min}= \! 1\,\msun$, no DM, 1 realization   \\ 
Asfe1    &   28.5 &   34.5 &    0.509 & $5.47\times10^5$ & 48.3 & 511 & 10.3     &1060 & 14.5  & $\alpha_{\rm sfe}= \! 1.26\times10^{-3}$, 2 realizations  \\ 
Asfe2    &  8.56 &   10.4  &   0.359 & $4.47\times10^5$  & 12.7 & 351 & 3.09   &  602  & 5.67 & $\alpha_{\rm sfe}= \! 3.16\times10^{-4}$ \\ 
Asfe3    &  1.75 &  2.10   &   0.361 & $8.51\times10^4$ & 8.61 & 144 & 0.629 &  186 & 1.67 & $\alpha_{\rm sfe}= \! 6.32\times10^{-5}$ \\ 
\hline

    \end{tabular}
    \label{tab:SC2}
  \end{center}
\end{table*}

It is important to examine how the initial configuration and
the values of our model parameters (see \secref{sec:generation})
affect the main result. 
To this end, we consider models with different parameters or
with different set-up:
models without no DM particles (AnoDM), adopting smaller stellar radii (Arad),
with different mass limits for the same IMF (Amax and Amin), and 
with different values of the SFE parameter $\alpha_{\rm sfe}$
(Asfe1, Asfe2, and Asfe3)
The assigned parameters and the resulting initial properties of the
star clusters are summarized in \tabref{tab:SC2}.

Let us first compare Model A and AnoDM. Clearly,
the existence of a dark matter halo has little impact on the
runaway growth of the stars. This is expected
because the characteristic time of dynamical friction between stars
and DM is very long.
We consider frictional force on a star with mass $m$ and velocity $v$
in a matter field 
with a density profile $\rho(r)$ \citep{Binney08}, and
use the equation of angular momentum change \citep{Fujii14},
to derive the dynamical friction time-scale $t_{\rm df}$
\begin{equation}
t_{\rm df}= \! \frac{0.186 v^3}{G^2m\ln\Lambda^\prime}\int^{r_{\rm i}}_0\frac{\dd r}{r\rho}, \label{eq:df}
\end{equation}
where $\Lambda^\prime\simeq \! 0.1N$ \citep{Giersz94} and $r_{\rm i}$ is
an initial radial position of the star.
Using the inner part of the Navarro-Frenk-White dark halo
density profile $\rho\propto r^{-1}$ 
with the normalization 
$\rho_0\sim \! 7\times10^{-20}\,{\rm g\,cm^{-3}}$
at $1\,\pc$ which is derived directly from our simulation, 
and using $v= \! 10\,{\rm km\,s^{-1}}$, $r_{\rm i}= \! 1\,\pc$, $m= \! 100\,\msun$
and $N= \! 3\times10^7$, we obtain
$t_{\rm df}\sim \! 7\,\Myr$ which is several dozen to hundreds times
larger than $t_{\rm cc}$. Hence the dynamical friction by dark matter 
is unimportant.

Next, we compare Model A and Arad. The difference is in the
model of stellar radius calculation. Model Arad assumes
a smaller stellar radius for a given stellar mass.
The difference causes only modest effect on the results.
Specifically, the number of collisions in Model Arad is
slightly smaller, but the final stellar mass $m_{\rm max,f}$ is almost the same.

A more direct effect is seen in Model Amax, where we set a
larger maximum mass limit for the power-law IMF $m_{\rm max}$.
The runaway growth of the central massive star is accelerated
in this case. Contrastingly, setting a smaller $m_{\rm min}$ (Model Amin)
does not significantly affect the runaway process.
Altogether, these results are reasonable because
mainly massive stars contribute to the runaway collisions
through mass segregation / concentration toward the cluster centre.
Note, however, that $m_{\rm min}$ is critically important in
determining the total stellar mass formed within a gas cloud,
hence the relative abundance of massive stars,
especially when one adopts a steep or bottom-heavy IMF. 
  
Varying the SFE parameter $\alpha_{\rm sfe}$ notably changes the results.
With a higher SFE (Model Asfe1), runaway collision is accelerated
because an initially denser cluster is made,
while setting a smaller $\alpha_{\rm sfe}$ results in smaller $m_{\rm max,f}$
and $N_{\rm coll}$ (Asfe2 and Asfe3).
In fact, in Model Asfe3, there are few collisions,  and the runaway collision
did not occur because of the small cluster mass
(note the small value of the second term in the equation \ref{eq:massev}).

\section{Summary and discussions}
\label{sec:summary}
We have performed a suite of simulations of star cluster evolution
and explored the formation of IMBHs through runaway collisions.
The star cluster formation sites are located in
realistic cosmological simulations.
Runaway stellar collisions occur quickly within $3\,\Myr$,
and very massive stars with mass of
$m_{\rm max,f}\sim \! 400-1900\,\msun$ (\figref{fig:mass_ind})
are formed, which is consistent with the results in \citet{Katz15}.
In all our fiducial models, $m_{\rm max,f}$ exceeds $\sim \! 300\,\msun$,
and thus IMBHs are expected to be left in the clusters \citep{Heger03}.
An interesting implication is that the IMBHs can seed the formation of  
the SMBHs observed at $z\gtrsim \! 6$.
The SMBHs are thought to have grown via gas accretion or merger.
If near-Eddington accretion is sustained with accretion
efficiency of $10\%$, 
a $1000\,\msun$ seed BH grows to become
as massive as $\sim \! 10^9\,\msun$ in $\sim \!  0.6\,\Gyr$ \citep[e.g.][]{Sakurai16}.
We conclude that IMBH formation in dense star clusters offers
a viable mechanism for seeding SMBHs.

\subsection{Correlation between the final mass and halo properties}
\begin{figure}
    \centering
    \includegraphics[width=0.5\textwidth]{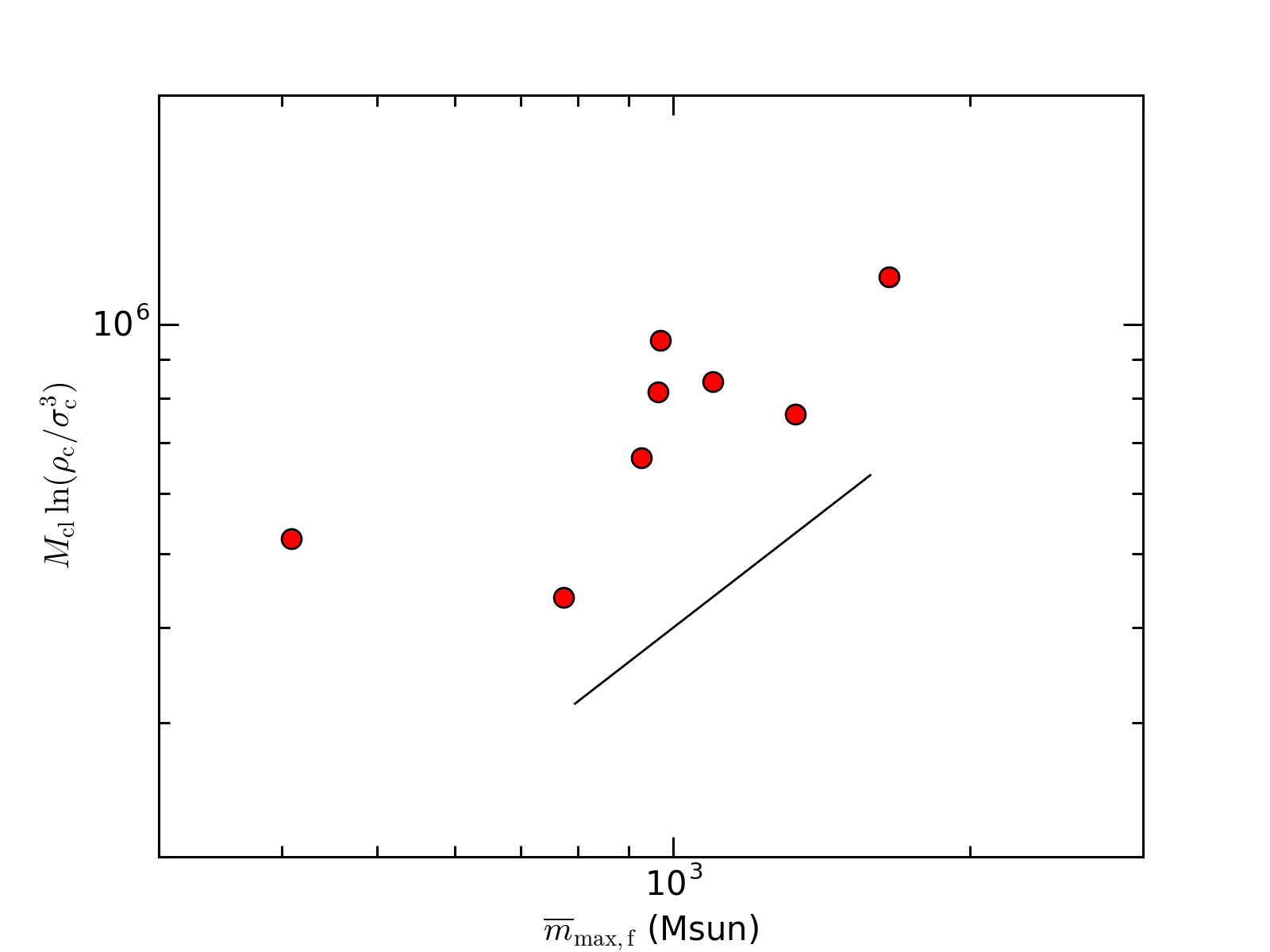}
    \caption{
    Correlation between $m_{\rm max,f}$ and $M_{\rm cl}\ln(\rho_{\rm c}/\sigma_{\rm c}^3)$, where $M_{\rm cl}$ is by $\msun$, $\sigma_{\rm c}$ is by ${\rm km\,s^{-1}}$, and $\rho_{\rm c}$ is by $\msun\,\pc^{-3}$. The line is $\propto m_{\rm max,f}$.
    }
    \label{fig:correlation}
\end{figure}

It is intriguing to study the origin of the diversity of
$m_{\rm max,f}$ (see \tabref{tab:SC}) found in our simulations.
To this end, we examine correlations between $m_{\rm max,f}$ and
several halo properties. Possibly relevant physical quantities are,
the halo virial mass $M_{\rm vir}$, central velocity dispersion of gas $\sigma_{\rm c,gas}$, and mean gas density of central core region $\overline{\rho}_{\rm c,gas}$.
Let us naively expect the cluster mass scales with the halo mass, 
$M_{\rm cl}\propto M_{\rm vir}$, and the stellar density scales with
the central gas density, $\rho_{\rm c}\propto\overline{\rho}_{\rm c,gas}$. 
From dynamical consideration, we also expect $\sigma_{\rm c}\propto\sigma_{\rm c,gas}$;
we assume essentially that the bulk properties of the star cluster are determined 
from those of the parent gas cloud. Then, because the core collapse time $t_{\rm cc}$
scales with the relaxation time $t_{\rm rc}$, 
equation \eqref{eq:massev} can be rewritten in terms of the halo properties as
\begin{equation}
  m_{\rm max,f}\propto M_{\rm vir}\ln({\rm const.}\times\overline{\rho}_{\rm c,gas}/\sigma_{\rm c,gas}^3)
  \label{eq:masshalo},
\end{equation}
with $t= \! 3\,\Myr$. The log dependence of $\sigma_{\rm c}$ and $\rho_{\rm c}$ comes
from integration
of the rate of mass growth by collision (scales as 
$\propto t^{-1}$) over a time interval of
$t_{\rm cc}\propto t_{\rm rc}\propto\sigma_{\rm c}^3/\rho_{\rm c}$ and $t= \! 3\,\Myr$ \citep{PortegiesZwart02}.
\figref{fig:correlation} shows the correlation between $m_{\rm max,f}$ and $M_{\rm cl}\ln(\rho_{\rm c}/\sigma_{\rm c}^3)$ 
for our fiducial models. Although with substantial scatter, we find
$m_{\rm max}$ is well correlated with the halo properties as in equation
\eqref{eq:masshalo}.

\begin{figure}
    \centering
    \includegraphics[width=0.5\textwidth]{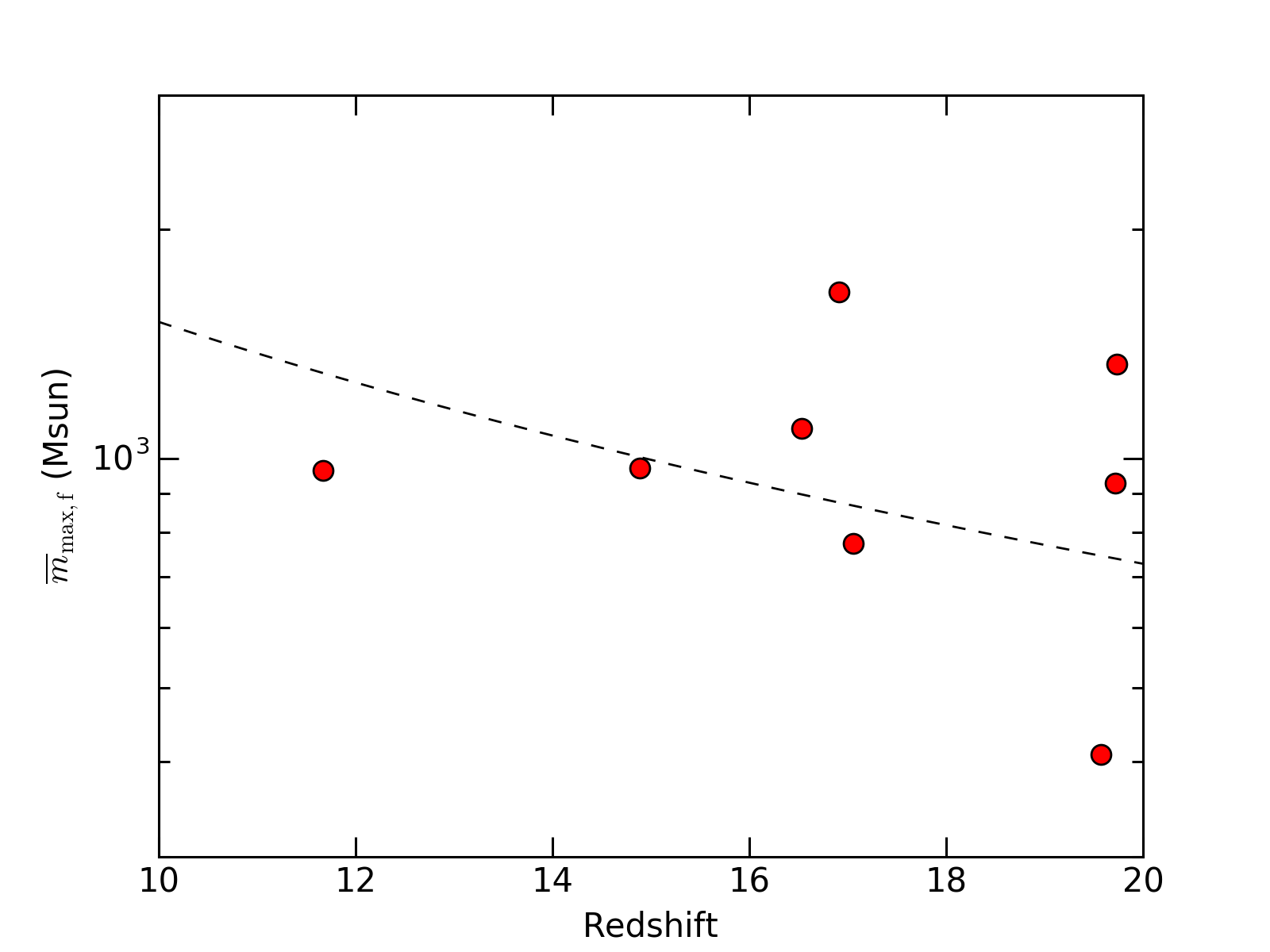}
    \caption{
    Correlation between redshift $z$ and $m_{\rm max,f}$. The dashed line is $\propto(1+z)^{-3/2}\ln(1+z)$ (see text).
    }
    \label{fig:correlation2}
\end{figure}

We can further derive a simple redshift dependence as follows.
We use the cosmological scaling of halo properties and redshift
as in, for example, the equations (18)-(20) of \citet{Ahn07};
$M_{\rm vir}\propto(1+z)^{-3/2}$ with constant $T_{\rm vir}\sim \! 8000\,{\rm K}$,
$\overline{\rho}_{\rm c,gas}\propto(1+z)^3$, 
and $\sigma_{\rm c,gas}\propto\sigma_{\rm gas}\sim \! (GM_{\rm vir}/r_{\rm t})^{1/2}\propto M_{\rm vir}^{1/3}(1+z)^{1/2}$. Then we obtain $m_{\rm max,f}\propto(1+z)^{-3/2}\ln(1+z)$.
\figref{fig:correlation2} shows  $m_{\rm max,f}$ against the cluster formation
redshift $z$. Again, we find that the actual dependence is consistent
with the above scaling (dashed line) but with substantial scatter.
Despite of these interesting correlations, we argue that
simulations with a significantly larger number of samples are needed to
determine more accurate correlations if there is any.

\subsection{Cluster mass-IMBH mass relation}
\begin{figure}
    \centering
    \includegraphics[width=0.5\textwidth]{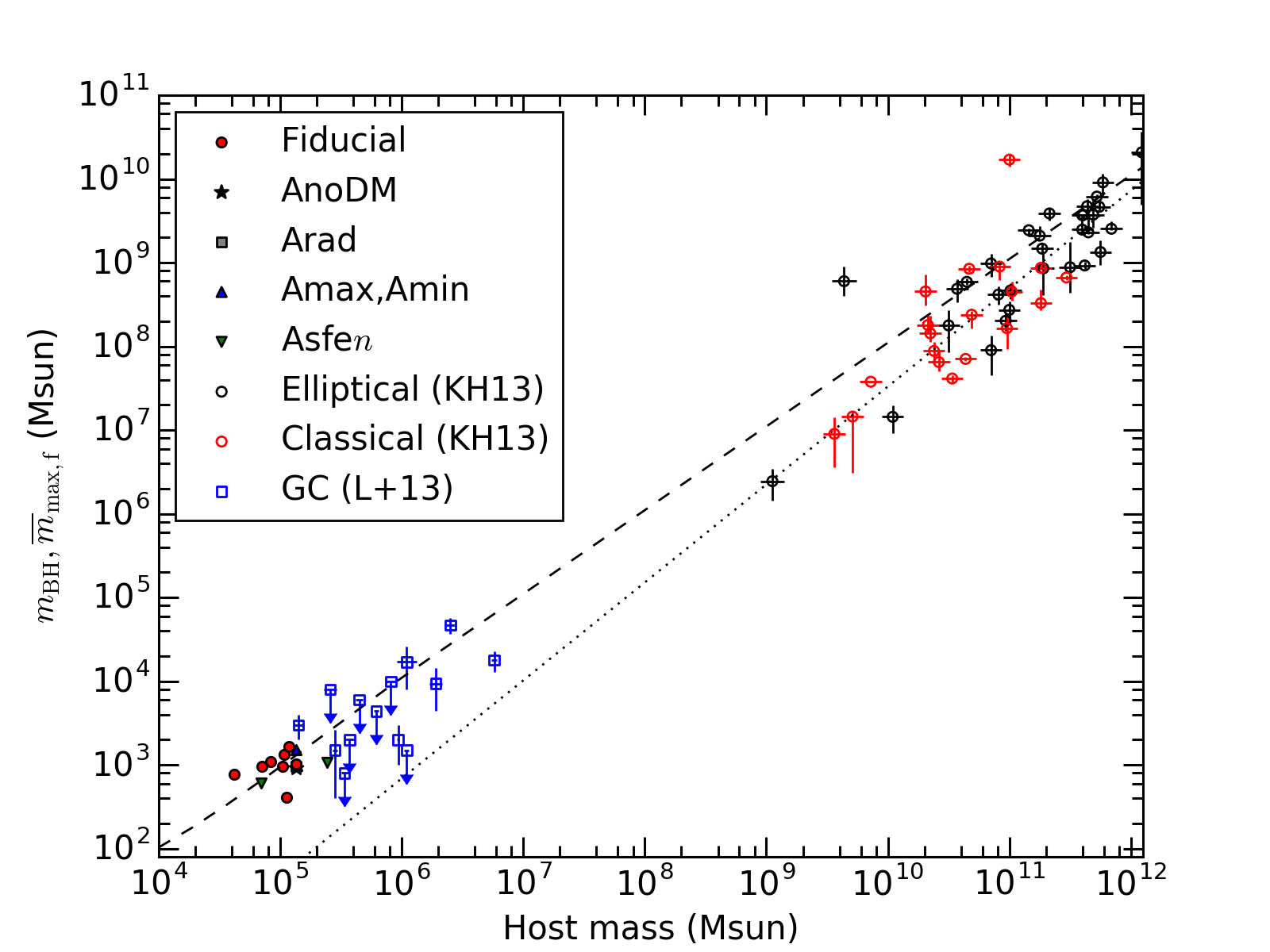}
    \caption{
    $\overline{M}_{\rm cl}$ vs. $\overline{m}_{\rm max,f}$ for the models in \tabref{tab:SC} and \tabref{tab:SC2}. The model Asfe3 is excluded since the massive star ends its life as a pair instability supernova with mass $\overline{m}_{\rm max,f}\lesssim \! 300\,\msun$ \citep{Heger03}.
    The observational data are also over-plotted for elliptical and classical bulges \citep{Kormendy13} and globular clusters \citep{Lutzgendorf13}.
    The dashed line and the dotted line are the analytical expressions of
    the black hole mass-cluster mass relation (equation \ref{eq:mass-cluster-rel}) and the 
    Magorrian relation (equation 10 of \citealt{Kormendy13}) respectively.
    }
    \label{fig:Mcl-mmax}
\end{figure}

There is a well-known SMBH mass-bulge mass relation 
\citep{Magorrian98,Merritt01} for galaxies and their central black holes.
Although the objects and mass scales are very different, it is
intriguing to compare the IMBH mass-cluster mass relation in
our simulations. 
In \figref{fig:Mcl-mmax}, we plot 
the final mass $\overline{m}_{\rm max,f}$ of the runaway collision stars
which are expected to collapse and leave IMBHs without significant mass loss
against cluster mass $\overline{M}_{\rm cl}$. Note that our samples are located
in the left-bottom portion of the figure.
The lower line indicates the
well-known Magorrian relation for SMBHs (equation 10 of \citealt[][]{Kormendy13}),
whereas the upper line is the black hole mass-cluster mass relation
\citep[see equation 16 and figure 3 of][]{PortegiesZwart02}.
More specifically, it is given by
\begin{equation}
m_{\rm max,f}= \! 30+8\times10^{-4}M_{\rm cl}\ln\Lambda_{\rm cl}, \label{eq:mass-cluster-rel}
\end{equation}
where $\Lambda_{\rm cl}= \! \min(M_{\rm cl}/\msun, 10^6)$.
Our results are indeed consistent with this. In other words, 
about one percent of the cluster mass contributes to the mass of the
central very massive star.
Overall, the first star clusters formed in early atomic-cooling haloes
may be quite similar to present-day star clusters, but
with smaller masses. The central black hole mass to galaxy (cluster) mass
ratio is slightly higher than that for local SMBHs, perhaps reflecting
the different formation mechanism.

\subsection{Model uncertainties}

The most significant uncertainty in our study lies in the process of generating
the initial conditions for star clusters (\secref{sec:generation}).
First of all, it is not trivial when a whole gas cloud
should be replaced by a star cluster. 
We assume that a cluster is formed
when the gas cloud density reaches $n_{\rm H}\sim \! 10^7\,{\rm cm^{-3}}$. 
We adopt the value by noting 
that efficient gas cooling and fragmentation by OH cooling occur 
in a gas cloud with metallicity $\gtrsim \! 10^{-4}\,\zsun$ \citep{Chiaki16}.
Although protostars are formed at much higher density of $\gtrsim \! 10^{20}\,{\rm cm^{-3}}$,
we set a lower density threshold for ``star cluster'' formation
to trace the global structure of the parent cloud.
Related to this assumption, we also note that
starburst is not strictly instantaneous but can last
for over a million years \citep{Kimm16}.
Continuous star formation may promote runaway collisions
by supplying newly formed stars to the cluster, or slower star formation
process can actually hamper the growth of the central star.
Also, a newly born star cluster does not necessarily have 
  the virial ratio $Q= \! 0.5$, and could have significantly anisotropic
  velocity structure.
Finally, the star formation efficiency 
can be regulated by stellar evolution itself
through various feedback effects, especially
when a different IMF is assumed. Self-consistent treatment of the
formation of star clusters and their evolution is beyond the scope of the
present paper, but direct cosmological simulations that couple star formation
and stellar dynamics as well as assembly of dark matter haloes will
clarify many of the above issues.

There are also several physical processes that are not included in our
hybrid $N$-body simulations. First, stellar evolution likely affects
the collision rate
through the evolution of the stellar radius and mass loss.
As already examined in Section 3.2, the overall uncertainty in the stellar radii
of main-sequence stars gives modest impact to our results. 
If post-main sequence evolution is considered, the collision rate
can be increased during the giant phases of the stars.
Stellar wind and mass loss are not expected to be
significant over a short time of $\sim \!  3\,\Myr$ if the stellar metallicity
is as low as $\lesssim \! 10^{-4}\,\zsun$
(\citealt{Baraffe01}; \citealt{Muijres12}; \citealt{Katz15}.
Second, binary evolution enhances mergers via tidal interaction
\citep{Hurley02}, and thus may promote runaway collisions. If this is the case,
the formation of super-massive stars can occur even faster than in our
simulations.
Third, primordial binaries can delay core collapse of clusters by binary
heating \citep{Rasio01}. They can also accelerate the core collapse by causing
mass segregation due to the effectively increased mass \citep{Heggie06}.
Finally, the actual collision conditions and outcome
may not be as simple as in our model (\secref{sec:Nbody}).
Tidal effects may enhance
the rate of close encounters (\citealt[][]{Fregeau04}, and references therein),
while mass loss during binary collision likely reduces the mass of
the collision product \citep[e.g.][]{Glebbeek09}.
Stellar rejuvenation by collisional mixing makes the lifetime of the collision
products longer.
Encountering two stars with high velocities that satisfy our merger condition 
could in reality just pass through each other depending on the impact parameter and the 
thickness of the stellar envelopes. Most of such stars in our simulations are
in tight binary orbits, and thus they may eventually merge at later time.
Interaction of stars and gas may enhance core collapse if dynamical friction
or mass augmentation by accretion is effective.
A dynamical friction time-scale of equation \eqref{eq:df} for gas is, assuming
an isothermal sphere profile $\rho= \! \rho_0(r/r_0)^{-2}$
with $\rho_0\sim \! 2\times10^{-18}\,{\rm g\,cm^{-3}}$ at $r_0= \! 1\,\pc$ derived from the halo data, and using $m= \! 100\,\msun$, 
$v= \! 10\,{\rm km\,s^{-1}}$, $r_{\rm i}= \! 1\,\pc$ and $\ln\Lambda^\prime\sim \! 10$, $t_{\rm df}\sim \! 0.2\,\Myr$.
An accretion time-scale is defined by $t_{\rm acc}= \! m/\dot{m}_{\rm B}$, 
where $m$ is a typical stellar mass and $\dot{m}_{\rm B}$ is the Bondi accretion rate \citep[e.g. ][]{Ryu16}.
For $m= \! 10\,\msun$, using temperature $\sim \! 10^4\,{\rm K}$ and density $n_{\rm H}\sim \! 10^7\,{\rm cm^{-3}}$ 
as indicated in \citet{Kimm16} for calculation of $\dot{m}_{\rm B}$, $t_{\rm acc}\sim \! 2\,\Myr$.
Both time-scales are within $3\,\Myr$ and the presence of gas can accelerate the core
collapse of the clusters.

\subsection{Fate of IMBHs in star clusters}
We have successfully shown the formation of super-massive stars in dense star clusters,
which undergo direct gravitational collapse to leave IMBHs. We thus suggest
a promising initial process of seeding the formation of SMBHs. 
An important question still remains, however.
It is unclear whether an IMBH in a star cluster can grow efficiently
to be a SMBH in about a billion years.
Either lack of gas supply or the BH's radiation feedback
can hamper its growth \citep[e.g.][]{Milosavljevic2009,Park_Ricotti2012}.
Some or even many IMBH might be left in star clusters or 
within galaxies in the present-day Universe
\citep[e.g.][]{Maccarone07,Pasham14,Kiziltan17}.
Further studies are warranted to examine the fate of the early IMBHs
and their observational signatures.

\section*{Acknowledgements}
YS is grateful to Junichiro Makino, Masaki Iwasawa, Kazumi Kashiyama, Tilman Hartwig, Kohei Inayoshi and Sunmyon Chon 
for fruitful discussions. 
The study is partly supported by Advanced Leading Graduate Course for Photon Science (YS),
by Grant-in-Aid for JSPS Research Fellows (15J08816: YS),
by JSPS KAKENHI Grant Number 26800108 (MSF), 
and by Grant-in Aid for JSPS Postdoctoral Fellowships for Research Abroad (SH).
The cosmological simulations and hybrid N-body simulations are performed on Cray XC30 at Center for Computational Astrophysics, National Astronomy Observatory of Japan and XC40 at YITP in Kyoto University.

\small{\bibliography{ms}}

\end{document}